\documentclass[twocolumn,amsmath,amssymb,aps,pra,superscriptaddress]{revtex4-2}

\usepackage{graphicx}
\usepackage{dcolumn}
\usepackage{bm}
\usepackage{color}
\newcommand{\Add}[1]{\textcolor{black}{#1}}

\begin{document}

\title{Edge-of-chaos enhanced quantum-inspired algorithm for combinatorial optimization}

\author{Hayato Goto}
\email[Corresponding author: ]{\Add{hayato.goto.d36@mail.toshiba}}
\affiliation{Corporate Laboratory, Toshiba Corporation, Kawasaki, Kanagawa 212-8582, Japan}
\affiliation{RIKEN Center for Quantum Computing (RQC), Wako, Saitama 351-0198, Japan}
\author{Ryo Hidaka}
\affiliation{Corporate Laboratory, Toshiba Corporation, Kawasaki, Kanagawa 212-8582, Japan}
\author{Kosuke Tatsumura}
\affiliation{Corporate Laboratory, Toshiba Corporation, Kawasaki, Kanagawa 212-8582, Japan}

\date{\today}

\begin{abstract}

Nonlinear dynamical systems with continuous variables can be used for solving combinatorial optimization problems with discrete variables. Numerical simulations of them are also useful as heuristic algorithms with a desirable property, namely, parallelizability, which allows us to execute them in a massively parallel manner, leading to ultrafast performance. However, the dynamical-system approaches with continuous variables are usually less accurate than conventional approaches with discrete variables such as simulated annealing. To improve the solution accuracy of a quantum-inspired algorithm called simulated bifurcation (SB), which was found from classical simulation of a quantum nonlinear oscillator network exhibiting quantum bifurcation, here we generalize it by introducing nonlinear control of individual bifurcation parameters and show that the generalized SB (GSB) can achieve surprisingly high performance, namely, almost 100\% success probabilities for some large-scale problems. As a result, the time to solution for a 2,000-variable problem is shortened to 10~ms by a GSB-based machine, which is two orders of magnitude shorter than the best known value, 1.3~s, previously obtained by an SB-based machine. To examine the reason for the ultrahigh performance, we investigated chaos in the GSB changing the nonlinear-control strength and found that the dramatic increase of success probabilities happens near the edge of chaos. That is, the GSB can find a solution with high probability by harnessing the edge of chaos. This finding suggests that dynamical-system approaches to combinatorial optimization will be enhanced by harnessing the edge of chaos, opening a broad possibility for physics-inspired approaches to combinatorial optimization.

\end{abstract}

\maketitle

\section{Introduction}

Selecting the best option among many candidates is important for making industrial and social activities more efficient. 
The mathematical formulation of this often results in a so-called combinatorial optimization problem with discrete variables such as bits. 
Unfortunately, such a problem is notoriously hard because of the exponentially large number of solution candidates with respect to the problem size, 
which is called combinatorial explosion~\cite{47}. 
To tackle such intractable problems, special-purpose machines~\cite{53}, 
especially so-called Ising machines~\cite{54}, have recently attracted much attention. 
A major approach to such machines is based on nonlinear dynamical 
systems~\cite{54,1,2,3,4,5,6,7,8,9,10,11,12,13,14,15,16,17,18,19,20,21,22,23,24,25,26,27,28,29,30,31,32,33,34,35,36,37,38,39,40,41,42,43,44,45}, 
such as neural networks~\cite{1,2}, nonlinear optical oscillator networks~\cite{3,4,5,6,7,8}, 
and nonlinear quantum oscillator networks~\cite{9,10,11,12,13,14}. 
In addition to real physical systems, 
numerical simulations of them are also useful as heuristic algorithms for combinatorial optimization~\cite{1,2,15,16,17,18,19,20,21,22,23,24,25,26,27,28,29,30,31,32}. 
Unlike classical systems such as neural networks and optical oscillator networks, quantum oscillator networks cannot be simulated efficiently, 
because they are enough to realize universal quantum computation~\cite{55,56}, 
which is believed not to be simulated efficiently by classical computers~\cite{57}. 
Thus, simulating a classical counterpart (classical Hamiltonian system~\cite{48,49,50}) of a quantum oscillator network exhibiting quantum bifurcation has been proposed, 
which is called simulated bifurcation (SB)~\cite{19,20,21,22,23,24,25}. 
Notably, the dynamical system-based algorithms usually have a desirable property: parallelizability. 
That is, we can execute the algorithms in a massively parallel manner, unlike a representative algorithm called simulated annealing (SA)~\cite{46,47}, 
leading to ultrafast performance by using cutting-edge many-core processors, such as graphics processing units (GPUs) and field-programmable gate arrays (FPGAs). 
However, the dynamical system-based algorithms with continuous variables are usually less accurate than conventional discrete-variable algorithms such as SA.

In this work, we focus on the SB and propose a new method for improving its solution accuracy. Because of its parallelizability and high solution accuracy, 
the SB has demonstrated ultrafast performance using GPUs and FPGAs~\cite{19,20,21,22,23}, and hence it is one of the most promising dynamical system-based algorithms. 
There are three major variants of SB: adiabatic SB (aSB) based on classical adiabatic evolution~\cite{19}, 
ballistic SB (bSB) obtained from aSB by replacing nonlinear potential walls with perfectly inelastic infinite walls~\cite{21}, 
and discrete SB (dSB) obtained from bSB by discretizing continuous variables via the sign function~\cite{21}. 
Note that all the SBs have a single bifurcation parameter controlling bifurcations of all the SB variables, 
which is usually scheduled deterministically, e.g., linearly with respect to time. 
The bSB can find good approximate solutions the fastest but cannot find best known solutions for large-scale problems~\cite{21}. 
The dSB was introduced to improve the accuracy of the bSB and succeeded in finding the best known solutions~\cite{21}.

Toward further improvement, here we generalize the bSB by introducing nonlinear control of individual bifurcation parameters corresponding to the bSB variables, 
which we call the generalized bSB (GbSB). Surprisingly, the GbSB can find the best known solutions with much higher probabilities, sometimes almost 100\%, 
for some large-scale problems than the dSB. Also importantly, the GbSB maintains the advantage of the SB, namely, parallelizability. 
To demonstrate it, we implemented the GbSB with an FPGA in a massively parallel manner and achieved one or two orders of magnitude shorter 
times to solution for large-scale problems than a previous dSB-based FPGA machine~\cite{21}. 
To examine the reason for the ultrahigh performance of the GbSB, we investigated chaos~\cite{51,52} in the GbSB changing the nonlinear-control strength. 
As is well known as the butterfly effect, the chaos causes a large difference between two trajectories with very close initial conditions due to nonlinearity. 
We actually observed that such two trajectories result in completely different states at the final time in the GbSB when the nonlinear-control strength is sufficiently large, 
clearly suggesting the chaos. 
Most importantly, we found that the dramatic increase of the success probabilities occurs near the edge of chaos in terms of the nonlinear-control strength. 
The edge-of-chaos effect has been known well in the field of artificial intelligence~\cite{58,59,60,61,62,63}, 
but not well for combinatorial optimization~\cite{64,65,66}. 
Thus, we expect that the SB enhanced by harnessing the edge of chaos will open a broad possibility for dynamical-system approaches to combinatorial optimization.

\section{Generalized ballistic simulated bifurcation (GbSB)}

In this work, we focus on the Ising problem, the target problem of Ising machines~\cite{54}, defined by the following objective (cost) function:
\begin{align}
E_{\mathrm{Ising}} =-\frac{1}{2} \sum_{i=1}^N \sum_{j=1}^N J_{i,j} s_i s_j,
\label{eq1}
\end{align}
where $s_i$ denotes the $i$th Ising spin taking 1 or $-1$, $N$ is the number of the spins, $J_{i,j}$ is the interaction coefficient 
between the $i$th and $j$th spins ($J_{i,j}=J_{j,i}$ and $J_{i,i}=0$), and $E_\mathrm{Ising}$ denotes the energy of the Ising model. 
The Ising problem is to find a spin configuration minimizing the Ising energy. 
This problem is known as an NP-hard problem~\cite{57,67}, and hence many other problems can be reduced to this problem with only polynomial overheads~\cite{68}.
To solve the Ising problem, the bSB numerically simulates the classical Hamiltonian dynamics with the following time-dependent Hamiltonian for $N$ oscillators:
\begin{align}
H_\mathrm{bSB} (t)=\frac{1}{2} \sum_{i=1}^N y_i^2 + \frac{p(t)}{2} \sum_{i=1}^N x_i^2 
- \frac{c}{2} \sum_{i=1}^N \sum_{j=1}^N J_{i,j} x_i x_j,
\label{eq2}
\end{align}
where $x_i$ and $y_i$ are the position and momentum, respectively, of the $i$th oscillator corresponding to the $i$th spin $s_i$ via $s_i=\mathrm{sgn}(x_i)$, 
$p(t)$ is a bifurcation parameter that varies from 1 to 0, and $c$ is a constant to tune the first bifurcation point~\cite{19,21}. 
To confine the positions within a finite range, we also introduce perfectly inelastic infinite walls at $x_i=\pm1$. 
More explicitly, the bSB is defined by the following update rules:
\begin{align}
&
y_i (t_{m+1} ) = y_i (t_m) - \frac{\partial H_{\mathrm{bSB}}}{\partial x_i} 
\Delta_\mathrm{t} 
\nonumber \\
&= y_i (t_m ) - \! \left[ 
p(t_{m+1} ) x_i (t_m ) - c \sum_{j=1}^N J_{i,j} x_j (t_m) \right] \! \! \Delta_\mathrm{t},
\label{eq3}
\\
&
x_i (t_{m+1} ) = x_i (t_m ) + \frac{\partial H_\mathrm{bSB}}{\partial y_i} 
\Delta_\mathrm{t} = x_i (t_m ) + y_i (t_{m+1} ) \Delta_\mathrm{t},
\label{eq4}
\\
&
\mathrm{If}~ |x_i (t_{k+1} )| > 1,~\mathrm{then}~ x_i (t_{k+1} ) \leftarrow 
\mathrm{sgn}[x_i (t_{m+1} )] 
\nonumber \\
& \qquad \qquad \qquad \qquad \mathrm{and}~ y_i (t_{m+1} ) = 0,
\label{eq5}
\end{align}
where $\Delta_\mathrm{t}$ is a time step and $t_m=\Delta_\mathrm{t} m$ ($m=0,1, \ldots, M$) is the discrete time ($t_M$ is the final time of the simulation). 
Note that the update rules given by Eqs.~(3) and (4) are based on the numerical integration method called the symplectic Euler method~\cite{48}, 
and Eq.~(5) describes the above-mentioned walls at $x_i=\pm 1$. 
Importantly, we can update all $x_i$ (all $y_i$) simultaneously according to Eq.~(4) [Eq.~(3)]. 
This is the parallelizability of the SB allowing for ultrafast performance~\cite{19,20,21,22,23}. 
[Note that the processing for the walls in Eq.~(5) is also parallelizable.]
The bSB has a single bifurcation parameter $p(t)$, which is usually scheduled linearly from 1 to 0 as ${p(t_m )=1-m/M}$ or equivalently
\begin{align}
p(t_{m+1} ) = p(t_m )-\frac{1}{M} = p(t_m ) - \frac{p(t_m )}{M-m}
\label{eq6}
\end{align}
with ${p(0)=1}$. 
We generalize the bSB by introducing individual bifurcation parameters $p_i (t)$ corresponding to 
$x_i$ [$p(t_{m+1} )$ in Eq.~(3) is replaced by $p_i (t_{m+1})]$ and controlling them by
\begin{align}
p_i (t_{m+1} ) = p_i (t_m ) - \! \left[ 1-Ax_i^2 (t_m ) \right] \! \frac{p_i (t_m )}{M-m},
\label{eq7}
\end{align}
where $A$ is a constant determining the strength of the nonlinear control of $p_i (t)$. 
When $A$ equals zero, the GbSB becomes the conventional bSB.

The meaning of Eq.~(7) is as follows. After a bifurcation, each oscillator approaches a wall, and consequently some of the oscillators stick to the walls in the middle of searching for a solution, 
resulting in trapping at a local minimum of the Hamiltonian (potential) with the walls. 
This is the reason why the conventional bSB usually results in only an approximate solution and cannot find an optimal one. 
To avoid sticking to the walls, the nonlinear control of $p_i (t)$ in Eq.~(7) reduces the decreasing rate of $p_i (t)$ for the oscillator close to a wall. 
Thus the individual bifurcation parameters are automatically controlled such that the oscillators do not stick to the walls, leading to higher solution accuracy 
by avoiding trapping at local minima. As shown below, this control leads to dramatic increase of the success probabilities of finding best known solutions for large-scale problems.

\begin{figure*}[ht]
	\includegraphics[width=15cm]{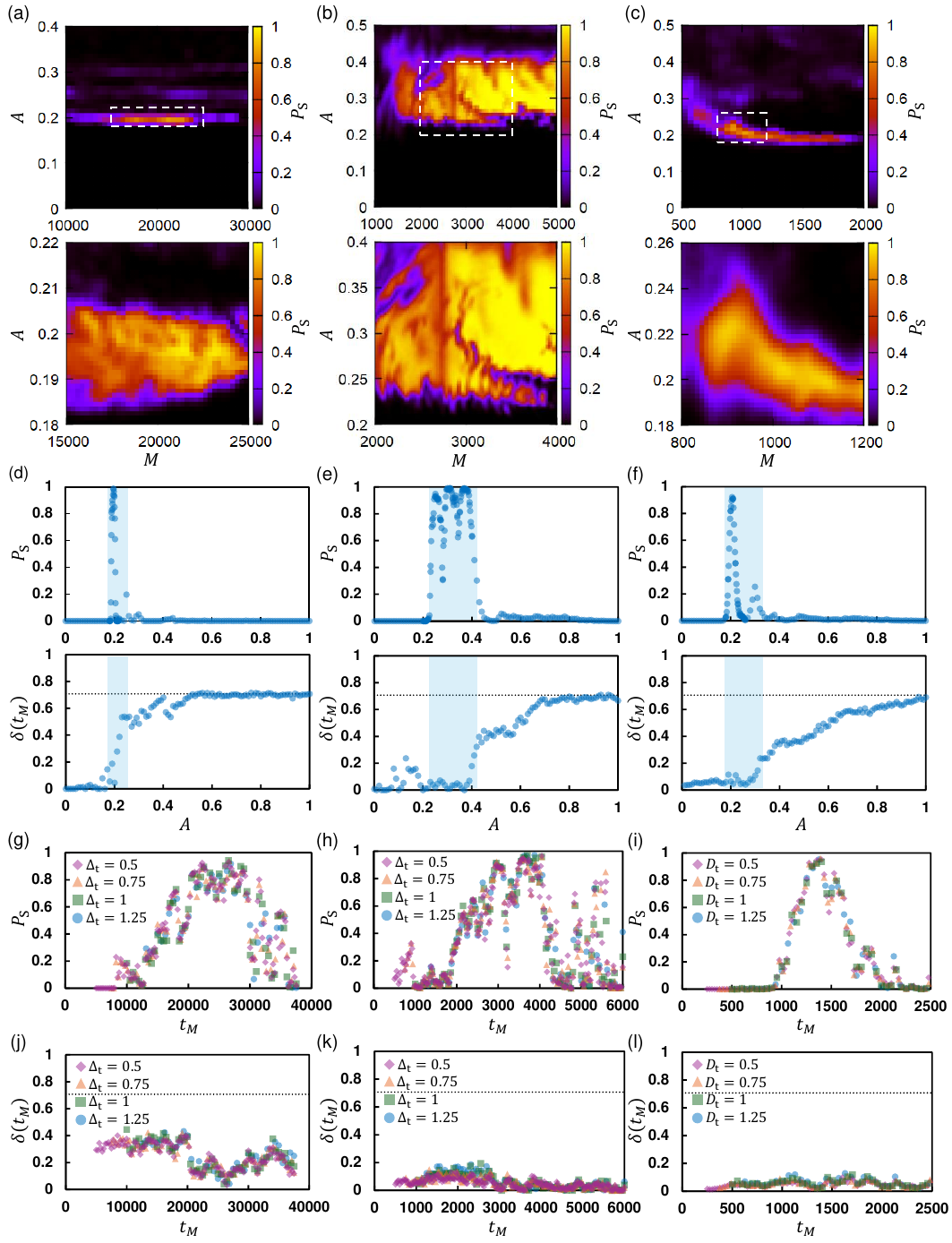}
	\caption{Numerical results of the GbSB. 
	(a) Success probability, $P_\mathrm{S}$, of finding the best known cut value of K$_{2000}$ by the GbSB with 
	${\Delta_t = 1.25}$ and various values of $M$ and $A$. 
	See Appendix~\ref{Appendix1} for the setting of $c$. 
	The lower figure is the enlargement of the part enclosed by the dashed rectangle in the upper figure. 
	$P_\mathrm{S}$ was evaluated by 1,000 times repetition with different initial conditions. 
	(b) and (c) Results corresponding to (a) for an instance of the 700-spin Ising problem and G6, respectively. 
	The time step is set as ${\Delta_t = 1.25}$ and ${D_t = 1.25}$ [see Eq.~(\ref{eq13}) in Appendix~\ref{Appendix1}], respectively.
	See Appendix~\ref{Appendix1} for the setting of $c$.
	(d) $P_\mathrm{S}$ in a (upper figure) and normalized distance, $\delta (t_M)$, at the final time (lower figure) for K$_{2000}$ when ${M=21,500}$ in (a). 
	$\delta (t_M)$ was evaluated by averaging 100 results with different initial conditions. 
	The regions highlighted in blue show the condition that $P_\mathrm{S}$ becomes particularly high. 
	The dotted line in the lower figure shows $\delta (t_M )=1/\sqrt{2}$, which indicates chaos. 
	(e) and (f) Results corresponding to (d) for the 700-spin instance when ${M=3,000}$ in (b) and 
	G6 when ${M=1,000}$ in (c), respectively. 
	(g--i) $P_\mathrm{S}$ for K$_{2000}$, the 700-spin instance, and G6, respectively, 
	for different values of the time step $\Delta_\mathrm{t}$ and different final times $t_M=\Delta_\mathrm{t} M$. 
	$A$ was set to 0.2, 0.25, and 0.2, respectively. 
	(j--l) $\delta (t_M)$ corresponding to (g--i), respectively.}
	\label{fig1}
\end{figure*}

\section{Performance of the GbSB}

To evaluate the performance of the GbSB, we first solved a standard benchmark problem named K$_{2000}$, 
a 2,000-node MAX-CUT problem~\cite{57} equivalent to a 2,000-spin Ising problem with all-to-all connectivity ($J_{i,j} \in \{ \pm 1 \}$)~\cite{5,17,19,21,24,32}. 
In this work, we evaluate the cut value only at the final time, because its calculation is time-consuming. 
(We confirmed that the present results remain unchanged even if we evaluate the cut value at all the time steps and select the best one.)
The best known cut value of K$_{2000}$ is 33,337 reported in ref.~\citenum{21}. 
Figure~1(a) shows the success probability, $P_\mathrm{S}$, of finding the best known value by the GbSB with various values of $M$ and $A$. 
[See Appendix~\ref{Appendix1} for the detailed settings of the other parameters. 
Also see Appendix~\ref{Appendix3} for average cut values corresponding to Figs.~1(a--c).] 
It thus turns out that the GbSB with appropriate values of $M$ and $A$ can find the best known value of K$_{2000}$ with almost 100\% probability. 
To the best of our knowledge, this is the highest success probability of this problem.
(In Fig.~\ref{fig1}(b), we can see a vertical stripe around ${M=2700}$ where $P_\mathrm{S}$ becomes relatively low. 
This is not an artifact but its underlying cause is difficult to explain. This may be a unique circumstance of this instance.)

To verify that the ultrahigh performance of the GbSB holds for problems other than K$_{2000}$, 
we also solved randomly generated 100 instances of the 700-spin Ising problem with all-to-all connectivity (${J_{i,j} \in \{ \pm 1 \}}$)~\cite{21,27} 
and the first ten instances of the well-known MAX-CUT benchmark set called G-set~\cite{5,15,16,17,21,26,28,30,31,32,45}, 
which are equivalent to the 800-spin Ising problem with sparse connectivity (${J_{i,j} \in \{ 0, -1 \}}$  or $\{0, \pm 1\}$, see Fig.~\ref{fig2}). 
Similarly to the case of K$_{2000}$, we obtained success probabilities exceeding 90\% for 32 instances of the 700-spin problem 
and two G-set instances (G3 and G6). (See Tables~\ref{table1} and \ref{table2} for the detailed results and settings.) 
Figures~\ref{fig1}(b)(c) show the corresponding results of two of them to Fig.~\ref{fig1}(a). 
Thus, the ultrahigh performance of the GbSB holds not only for K$_{2000}$ but also for other large-scale problems.

\section{Edge-of-chaos effect in the GbSB}

To examine the reason for the ultrahigh performance of the GbSB, 
we investigated chaos in the GbSB as follows. 
We run the GbSB with two sets of initial conditions: ${x_i^{(1)} (0)=\pm 0.1}$ ($\pm$ is randomly chosen) 
and ${y_i^{(1)} (0)=0}$; ${x_i^{(2)} (0)=x_i^{(1)} (0)\pm 2\times 10^{-6}}$ 
($\pm$ is randomly chosen) and ${y_i^{(2)} (0)=0}$. 
Note that the two trajectories are very close to each other at the initial time. 
Next, we introduce the normalized distance ${\delta (t)}$ between the two trajectories as
\begin{align}
\delta (t_m ) = \sqrt{\frac{1}{4N} \sum_{i=1}^N \left[ x_i^{(1)} (t_m ) - x_i^{(2)} (t_m ) \right]^2}.
\label{eq8}
\end{align}
From the above initial conditions, ${\delta (0)=10^{-6}}$. 
Also note that ${\delta \le 1}$ from ${|x_i^{(1)} | \le 1}$ and ${|x_i^{(2)} | \le 1}$, 
and the equality holds when ${|x_i^{(1)} - x_i^{(2)} | = 2}$ for all $i$. 
Most importantly, ${\delta (t) \simeq 1/\sqrt{2}}$ when all $x_i^{(1)}$ and $x_i^{(2)}$ become $\pm 1$ randomly. 
That is, ${\delta (t) \simeq 1/\sqrt{2}}$ indicates the chaos.

The lower figures in Figs.~\ref{fig1}(d--f) show the results of $\delta (t_M )$ ($\delta$ at the final time) corresponding to Figs.~\ref{fig1}(a--c), respectively. 
In all the three cases, ${\delta (t_M ) \simeq 0}$ when ${A=0}$, indicating regular dynamics, 
and $\delta (t_M )$ converges to ${1/\sqrt{2}}$ (dotted lines) as $A$ increases, 
indicating the chaos due to the nonlinear control. 
The regions highlighted in blue in Figs.~\ref{fig1}(d--f) show the condition that the success probabilities become particularly high, 
as shown in the upper figures. 
We thus find that the success probabilities become high near the edge of chaos (between regular and chaotic regions). 
This result suggests that weakly chaotic processes occurring near regular dynamics may assist avoiding trapping at local minima and finding optimal solutions in the GbSB. 
This chaos-assisted solution search may be essentially different from conventional approaches with random noises such as SA, 
because the dramatic increase of the success probabilities, in particular, almost 100\% success probabilities for the large-scale problems, has not been achieved by SA.

\section{No influence of discretization}

We also investigated the dependence of the performance on the time step $\Delta_\mathrm{t}$. 
The results in Figs.~\ref{fig1}(g--i) show that the notable behavior of the GbSB is independent of $\Delta_\mathrm{t}$. 
This result suggests that the high performance of the GbSB harnessing the edge of chaos comes not from artifacts due to the discretization for numerical simulations, 
but from the original Hamiltonian equations of motion with the walls and the nonlinear control of individual bifurcation parameters.

Figures~\ref{fig1}(g--i) also show that $P_S$ does not monotonically increase with respect to the number of steps $M$, unlike SA.
To investigate the reason for this, we checked the normalized distance at the final time $\delta (t_M )$ corresponding to Figs.~\ref{fig1}(g--i), 
the results of which are shown in Figs.~\ref{fig1}(j--l).
Interestingly, Fig.~\ref{fig1}(j) suggests that $P_S$ increases around the ``valley of chaos.''
This is also the case in Figs.~\ref{fig1}(k) and \ref{fig1}(l), 
though they are less clear. 
\Add{Note that a longer $t_M$ (larger $M$) leads to lower decreasing rates of the bifurcation parameters through Eq.~(\ref{eq7}). 
This change of the decreasing rates may affect the chaos in the present system and lead to the valley of chaos.
So far, however, it is difficult to understand how the success probability $P_S$ increases through this change, 
which is left for future work.}

\clearpage

\section{GbSB-based machine with an FPGA}

\begin{figure*}[ht]
	\includegraphics[width=13cm]{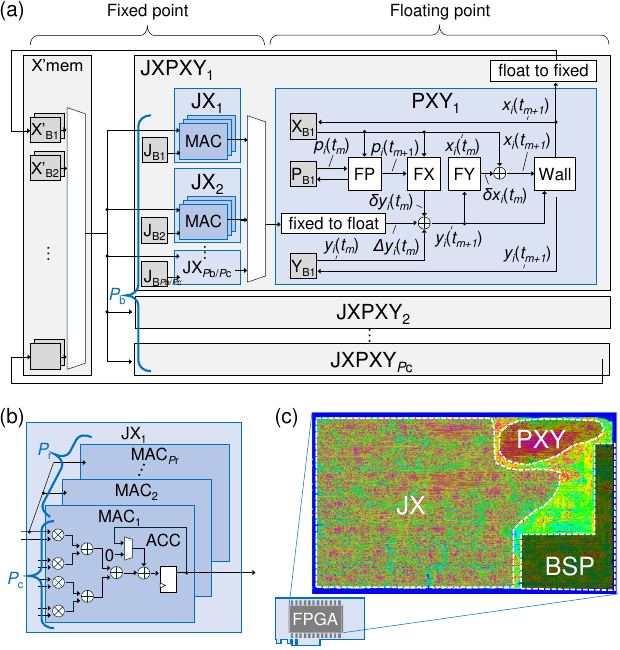}
	\caption{GbSB-based machine with an FPGA. 
	(a) Block diagram of the GbSB-based machine, which has a circulative structure corresponding to iteration of updating position and momentum: 
	$y_i (t_{m+1} )=y_i (t_m )+\delta y_i (t_m ) + \Delta y_i (t_m )$ and 
	$x_i (t_{m+1} )=x_i (t_m )+\delta x_i (t_m )$, 
	where $\Delta y_i (t_m ) = \mathrm{JX}[\mathbf{J}_i, \mathbf{x}(t_m)] = c \Delta_\mathrm{t} \sum_{j=1}^N J_{i,j} x_j (t_m)$, 
	$p_i (t_{m+1} )=\mathrm{FP}[p_i (t_m ), x_i (t_m )] = p_i (t_m ) - [1-A x_i (t_m)^2 ] p_i (t_m )/(M-m)$, 
	$\delta y_i (t_m ) = \mathrm{FX}[p_i (t_{m+1} ), x_i (t_m )] = -p_i (t_{m+1} ) x_i (t_m ) \Delta_\mathrm{t}$, and 
	$\delta x_i (t_m ) = \mathrm{FY}[y_i (t_{m+1} )] = y_i (t_{m+1} ) \Delta_\mathrm{t}$. 
	The JX, FP, FX, FY modules in the block diagram correspond to the JX, FP, FX, FY functions, 
	and the wall module executes the following conditional operation: 
	$(x_i (t_{m+1} ),y_i (t_{m+1} )) \leftarrow \left\{ 
	\begin{array}{ll}
	(\mathrm{sgn} (x_i (t_{m+1} )), 0), & \mathrm{if}~|x_i (t_{m+1} )|>1 \\ 
	(x_i (t_{m+1} ), y_i (t_{m+1} )), & \mathrm{otherwise}
	\end{array} \right.$. 
	The computational precision of the JX module is 16-bit fixed point, while that for the remaining modules is 32-bit floating point. 
	(b) Block diagram of a many-body interaction (JX) module. 
	The parallelization parameters ($P_\mathrm{r}$, $P_\mathrm{c}$, and $P_\mathrm{b}$) are illustrated in \textbf{a} and \textbf{b}. 
	(c) Layout of the circuit modules in the FPGA, where the routing congestion is shown 
	as a heatmap with the regions for the JX, PXY, and BSP (board support package) modules being indicated by dashed lines.}
	\label{fig2}
\end{figure*}

\begin{figure*}[ht]
	\includegraphics[width=15cm]{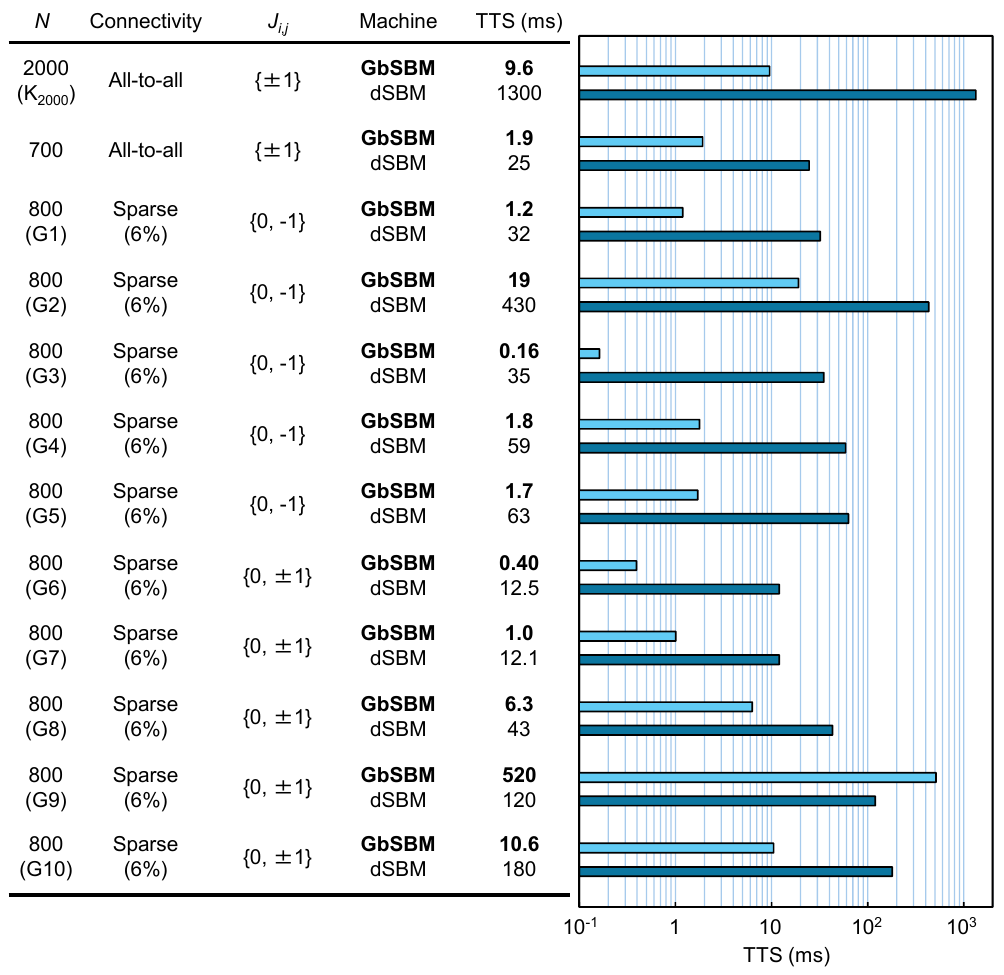}
	\caption{Times to solution (TTSs) with a GbSB-based FPGA machine. 
	The second row (${N=700}$) is the median of the TTSs for 100 random instances of the 700-spin Ising problem. 
	See Appendix~\ref{Appendix1} for the detailed parameter settings and the GbSB-based machine (GbSBM). 
	The results with the dSB-based machine (dSBM) are cited from ref.~\citenum{21}.}
	\label{fig3}
\end{figure*}

To demonstrate high parallelizability of the GbSB, 
we designed a highly parallel custom circuit for the GbSB to minimize the execution time and then instantiated a 2,048-spin GbSB machine (GbSBM) with an FPGA (Fig.~\ref{fig2}). 
Each individual bifurcation parameter $p_i (t)$ newly introduced in the GbSB is updated depending only on its corresponding position $x_i (t)$ without depending on the other oscillators. 
Hence, the GbSB is highly parallelizable, as conventional SBs are~\cite{19,20,21,22,23}. 
The increase in computational complexity resulting from introducing individual parameters ${p_i (t)}$ [Eq.~(\ref{eq7})] 
instead of a common parameter ${p(t)}$ [Eq.~(\ref{eq6})] is ${\Theta (N)}$.
In the computation for a GbSB time-evolution step, the many-body interaction computation, 
$\sum_{j=1}^N J_{i,j} x_j (t_m )$, in Eq.~(3) is the most computationally intensive with a computational complexity of $\Theta (N^2)$, 
while the remaining parts including updating $p_i (t)$, $x_i (t)$, and $y_i (t)$, correctively referred to as PXY, have a computational complexity of $\Theta (N)$. 
We accelerate the many-body interaction computation by spatial parallelization [referred to as JX in Fig.~\ref{fig2}(a)], 
which is characterized by three parallelization parameters of $P_\mathrm{r}$, $P_\mathrm{c}$, and $P_\mathrm{b}$ [Figs.~\ref{fig2}(a)(b)]. 
The PXY part is accelerated by temporal parallelization [Fig.~\ref{fig2}(a)]. 
The number of the clock cycles per GbSB time-evolution step is expressed as
\begin{align}
N_\mathrm{cyc} = \frac{N^2}{P_\mathrm{r} P_\mathrm{c} P_\mathrm{b}} + \frac{P_\mathrm{b} P_\mathrm{r}}{P_\mathrm{c}} + \lambda,
\label{eq9}
\end{align}
where $\lambda$ is the maximum circuit latency for circulative paths (See Appendix~\ref{Appendix2} for details). 
The circuit architecture accelerates the many-body interaction computation by $P_\mathrm{r} P_\mathrm{c} P_\mathrm{b}$ and 
allows that $P_\mathrm{r} P_\mathrm{c} P_\mathrm{b}$ can be larger than the number of oscillators $N$.

Dynamical system-based algorithms for combinatorial optimization often require complex functions, such as trigonometric functions~\cite{30} 
and hyperbolic tangent~\cite{1,2,29} (or other nonlinear functions~\cite{26}), 
and also pseudorandom number generators~\cite{17,29}. 
Such functions consume a lot of computation resources when implemented with FPGAs. 
Unlike those, the GbSB is based on only simple computations such as addition and multiplication, 
enabling a massively parallel implementation fully utilizing the capability of FPGAs. 
Figure~\ref{fig2}(c) shows the circuit layout of the GbSBM (${N=}$2,048) in an FPGA. 
The computation parallelism, the number of multiply accumulation (MAC) processing elements for many-body interaction computation, 
achieved in this work is $P_\mathrm{r} P_\mathrm{c} P_\mathrm{b}=32,768$, where $P_\mathrm{r}$, $P_\mathrm{c}$, and $P_\mathrm{b}$ are 8, 32, and 128, respectively, 
leading to $N_\mathrm{cyc}$ of 260 clock cycles ($\lambda$ is 100 clock cycles). 
We fully utilized the computing resources of the FPGA (the utilization of logic elements is 94\%) 
and concurrently achieved almost the highest system clock frequency achievable with the FPGA (${F_\mathrm{sys}=591}$~MHz). 
Consequently, the time per GbSB time-evolution step is 0.440~$\mu$s. 
See Table~\ref{table3} for details.

The GbSBM is evaluated in terms of times to solution (TTSs) for the above problems. 
The TTS is defined as the computation time required for finding an optimal or best known solution with 99\% probability. 
Mathematically, the TTS is formulated as $T_\mathrm{com} {\log (1-0.99)}/ {\log (1-P_\mathrm{S})}$, 
where $T_\mathrm{com}$ is the computation time of a single run and if ${P_\mathrm{S}>0.99}$, 
the TTS is defined by $T_\mathrm{com}$. 
The statistical error of the TTS, $\Delta_\mathrm{TTS}$, is given by 
$\Delta_\mathrm{TTS}=\mathrm{TTS} \times \Delta P_\mathrm{S}/[(1-P_\mathrm{S})|\ln (1-P_\mathrm{S} ) |]$, 
where ${\Delta P_\mathrm{S}}$ is the statistical error of $P_S$ and formulated as 
$\Delta P_\mathrm{S}=\sqrt{(P_\mathrm{S}-P_\mathrm{S}^2 )/N_\mathrm{rep}}$ ($N_\mathrm{rep}$ is the number of repetitions for estimating $P_\mathrm{S}$)~\cite{21}. 
The 2,048-spin GbSBM can solve the 700-spin instances and the 800-spin G-set instances twice at the same time~\cite{21}. 
We did such a batch processing for these instances (the batch number is denoted by $N_\mathrm{batch}$) and chose a better result, 
which enhances $P_\mathrm{S}$ while keeping the same $T_\mathrm{com}$.

The results of the TTSs are summarized and compared with those obtained by a previously developed FPGA-based dSB machine (dSBM)~\cite{21} in Fig.~\ref{fig3}, 
where dSB is obtained from bSB by replacing $J_{i,j} x_j (t_m)$ with $J_{i,j} s_j (t_m)$ in Eq.~(3). 
First, the TTS for K$_{2000}$ is shortened to 9.6~ms by the GbSBM from 1.3~s of the dSBM, 
leading to two orders of magnitude improvement. 
Similarly, the median of the TTSs for the 100 instances of the 700-spin Ising problem and the TTSs for the G-set instances are one or two orders of magnitude shortened except for G9. 
These results support the expectation that the GbSB harnessing the edge of chaos will substantially enhance the performance of SB-based machines.

\section{Summary and outlook}
We have generalized bSB by introducing nonlinear control of individual bifurcation parameters such that the oscillators in the bSB do not stick to walls. 
We have found that the success probabilities of the generalized bSB (GbSB) for large-scale problems dramatically increase and sometimes approach almost 100\%. 
Developing a GbSB-based machine with an FPGA, we have demonstrated one or two orders of magnitude improvements of times to solution for the large-scale problems 
compared with a previously developed SB-based FPGA machine. 
By investigating chaos in the GbSB, we have also found that the surprising increase of the success probabilities occurs near the edge of chaos. 
This result suggests that the GbSB can achieve the ultrahigh performance by harnessing the edge of chaos. 
The present study on chaos is based on numerical experiments. 
Theoretical studies on the present system are desirable but left for future work.
Also, the above conclusion is based on numerical study on 111 specific instances, and therefore 
investigating more instances from a wider range of problem classes is also desirable.

Our results open broad and interesting possibilities as follows. 
In this work, we have focused on bSB and found the edge-of-chaos effect. 
It is thus interesting whether or not other dynamical-system approaches, such as aSB~\cite{19}, dSB~\cite{21}, neural networks~\cite{1,2}, 
coherent Ising machines~\cite{3,4,5,6,7,8}, and oscillator-based Ising machines~\cite{30,31}, 
exhibit similar effects by similar nonlinear control. 
The GbSB has an additional parameter, namely, the nonlinear-control strength, 
which requires additional tuning for achieving high performance, 
together with tuning of the number of time steps. 
Although we have not included the time to optimize these parameters in computation times in this work, as in previous studies, 
the optimization time is not negligible for practical applications. 
\Add{So far, the parameter optimization relies on grid search, as shown in Figs.~\ref{fig1}(a--c).}
To optimize the parameters \Add{more} quickly, 
our finding that good parameter values may be found around the edge of chaos 
will be helpful.
The GbSB could not improve G9, one of the G-set instances, compared with dSB. 
This result suggests the limitation of the power of GbSB. 
While it is important to clarify the range of applicability of GbSB, 
this result suggests that we may need a hybrid algorithm combining GbSB with others for solving a wide range of problems.
In general, the SBs can achieve high performance for dense-connectivity problems~\cite{54,69}. 
On the other hand, a dynamical-system approach that can achieve ultrahigh performance for sparse-connectivity problems has recently been reported~\cite{70}. 
The combination of them will be promising to tackle both kinds of problems.

\section*{Acknowledgements}
RH and KT thank Masaya Yamasaki for his kind help.

\begin{appendix}

\section{Parameter settings in the GbSB}
\label{Appendix1}

The GbSB has two parameters: a time step $\Delta_\mathrm{t}$ and a constant $c$ for tuning the first bifurcation point, 
similarly to the other SB algorithms~\cite{19,20,21,22,23,24,25}. 
In this work, we set them as follows. 
In the cases of conventional SB algorithms with a single bifurcation parameter $p(t)$, 
the first bifurcation point is given by ${p(t)=c \lambda_\mathrm{max}}$, where $\lambda_\mathrm{max}$ is the largest eigenvalue of the interaction matrix $J$~\cite{19,21}. 
$\lambda_\mathrm{max}$ is positive because the trace of $J$ is zero. 
To set the first bifurcation point at the initial time for eliminating useless dynamics before the bifurcation, 
$c$ is usually set as ${c=p(0)/\lambda_\mathrm{max}=1/\lambda_\mathrm{max}}$~\cite{19,21}. 
Similarly, we set ${c=1/\lambda_\mathrm{max}}$ in the GbSB because ${p_i (0)=1}$. 
Note that we can estimate $\lambda_\mathrm{max}$ for a random matrix, 
such as that of K$_{2000}$, as ${\lambda_\mathrm{max}=2\sqrt{N} \sigma}$ from Wigner’s semicircle law, 
where $\sigma$ is the standard deviation of the nondiagonal elements of $J$. 
In this work, we thus set $c=1/(2\sqrt{N} \sigma)=1/(2\sqrt{N})$ for K$_{2000}$ and 
the 100 instances of the 700-spin Ising problem. 
For the ten instances of G-set, we numerically evaluate $\lambda_\mathrm{max}$ and set ${c=1/\lambda_\mathrm{max}}$.

For the setting of $\Delta_\mathrm{t}$, 
we consider the stability condition for the Hamiltonian dynamics. 
At the initial time, the Hamiltonian system is equivalent to the system of $N$ harmonic oscillators with mass of unity and spring constants of 
${k_i=p(0)-c \lambda_i=1-\lambda_i/\lambda_\mathrm{max}}$, where $\lambda_i$ is the $i$th eigenvalue of $J$. 
The position and momentum of the $i$th harmonic oscillator are denoted by $x_i$ and $y_i$, respectively. 
Then, from the update rules of the symplectic Euler method~\cite{48} in Eqs.~(3) and (4), 
we have $y_i (t_{m+1} )=y_i (t_m )-k_i x_i (t_m ) \Delta_\mathrm{t}$ and 
$x_i (t_{m+1} )=x(t_m )+y_i (t_{m+1}) \Delta_\mathrm{t}$, which result in
\begin{align}
x_i (t_{m+1} )-(2-k_i \Delta_\mathrm{t}^2 ) x_i (t_m ) + x_i (t_{m-1} )=0.
\label{eq10}
\end{align}
The condition that Eq.~(\ref{eq10}) describes oscillation is given by $(2-k_i \Delta_\mathrm{t}^2 )^2-4<0$, 
leading to the following stability condition for the $i$th harmonic oscillator:
\begin{align}
\Delta_\mathrm{t} < \frac{2}{\sqrt{k_i}} 
= \frac{2}{\sqrt{1-\lambda_i/\lambda_\mathrm{max}}}.
\label{eq11}
\end{align}
Thus, the sufficient condition for $\Delta_\mathrm{t}$ satisfying Eq.~(\ref{eq11}) for all the harmonic oscillators is given by
\begin{align}
\Delta_\mathrm{t} < 
\frac{2}{\sqrt{1-\lambda_\mathrm{min}/\lambda_\mathrm{max}}},
\label{eq12}
\end{align}
where $\lambda_\mathrm{min}$ is the minimum eigenvalue of $J$. 
Note that $\lambda_\mathrm{min}$ is negative, because the trace of $J$ is zero, 
and can be estimated as ${\lambda_\mathrm{min}=-2\sqrt{N} \sigma}$ for a random matrix. 
In the case of a random matrix, Eq.~(\ref{eq12}) is reduced to ${\Delta_\mathrm{t} < \sqrt{2} = 1.414\cdots}$. 
This is the reason why ${\Delta_\mathrm{t}=1.25}$ resulted in good performance for K$_{2000}$ and other random instances in the previous study~\cite{21}. 
In this work, we also used ${\Delta_\mathrm{t}=1.25}$ for K$_{2000}$ and the 100 instances of the 700-spin Ising problem. 
We can generalize this setting to the general case by using Eq.~(\ref{eq12}) as follows:
\begin{align}
\Delta_\mathrm{t}=D_\mathrm{t} \sqrt{\frac{2}{1-\lambda_\mathrm{min}/\lambda_\mathrm{max}}}
\label{eq13}
\end{align}
with ${D_\mathrm{t}=1.25}$.
We used this setting of $\Delta_\mathrm{t}$ for the ten instances of G-set in this work. 

As for initial conditions, $x_i (0)$ was set to a random value between $-0.1$ and 0.1, 
except the simulations to evaluate $\delta (t_M )$ in Figs.~\ref{fig1}(d--f), and $y_i (0)$ was set to zero.

\section{Average cut values in Figs.~\ref{fig1}(a--c)}
\label{Appendix3}

\begin{figure*}[h]
	\includegraphics[width=15cm]{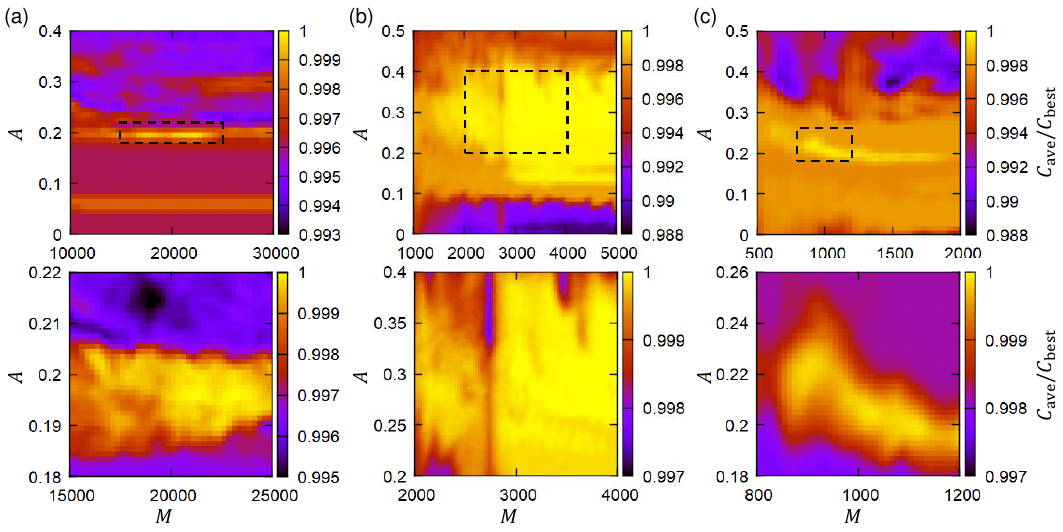}
	\caption{Average cut value $C_{\mathrm{ave}}$ normalized by the best know value $C_{\mathrm{best}}$ corresponding to Figs.~\ref{fig1}(a--c), respectively.}
	\label{fig5}
\end{figure*}

Figures~\ref{fig1}(a--c) show the success probability $P_S$ to obtain the best known value, 
where $P_S \simeq 0$ in a wide range of the parameters. 
Here we show the average cut value $C_{\mathrm{ave}}$ normalized by the best know value $C_{\mathrm{best}}$ 
in Fig.~\ref{fig5}.
We can confirm the gradual improvement of  $C_{\mathrm{ave}}$ by GbSB.

\section{FPGA implementation of GbSB}
\label{Appendix2}

\begin{figure*}[h]
	\includegraphics[width=12cm]{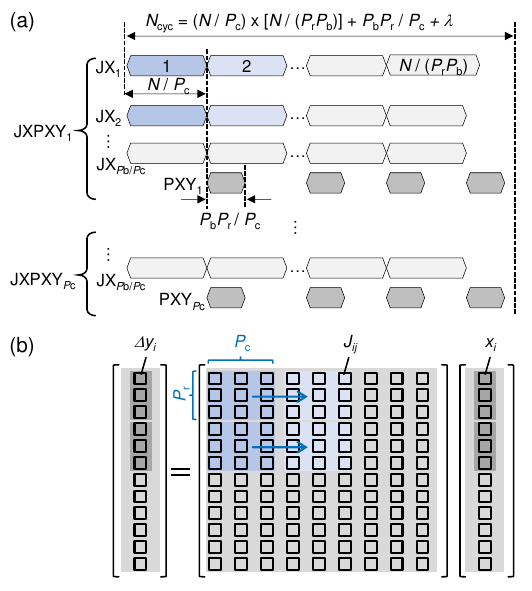}
	\caption{Parallel processing of the computation in a GbSB time-evolution step. 
	(a) Timing chart of the operation of JX and PXY modules. 
	(b) A schematic showing how the many-body interaction computations, 
	$\sum_{j=1}^N J_{i,j} x_j (t_m)$, are processed in parallel.}
	\label{fig4}
\end{figure*}

Figure~\ref{fig4}(a) shows the timing chart of the operation of JX and PXY modules. 
As shown in Fig.~\ref{fig4}(b), a JX module processes $P_\mathrm{r} P_\mathrm{c}$ multiply-accumulation operations per clock cycle and 
then produce $P_\mathrm{r}$ numbers of $\Delta y_i$ after $N/P_\mathrm{c}$ cycles (hereafter, a phase), 
and there are $P_\mathrm{b}$ JX modules. 
A PXY module shown in Fig.~\ref{fig2}a takes as input a $\Delta y_i$ datum per clock cycle and processes $P_\mathrm{b} P_\mathrm{r}/P_\mathrm{c}$ 
numbers of $\Delta y_i$ data per phase, and there are $P_\mathrm{c}$ PXY modules. 
The operation of a PXY module is overlapped with that of JX modules in time domain [Fig.~\ref{fig4}(a)]. 
The maximum circuit latency for circulative paths is denoted by $\lambda$, 
meaning that the system must wait for $\lambda$ to guarantee that the final datum has been stored in memory after the final iteration begins. 
Thus, the number of the clock cycles per GbSB time-evolution step, $N_\mathrm{cyc}$, is expressed by Eq.~(9).

Table~\ref{table3} summarizes the details of implementation and performance of the FPGA-based GbSB machine. 
The computational precision of the PXY module is 32-bit floating point, while that of the JX module uses less precision, i.e., 16-bit fixed point, to increase the parallelism of JX modules. 
The FPGA board used in this work (Bittware IA-840F Intel Agilex FPGA card) is equipped with an Intel Agilex7F AGF027 FPGA. 
The design is described and synthesized using Quartus Prime Pro 21.4 CAD tool. 
The limiting resource to determine the achievable maximum parallelism is logic elements used to synthesize JX modules. 
The system clock frequency $F_\mathrm{sys}$ determined as a result of circuit synthesis, placement, 
and routing is 591~MHz, corresponding to clock cycle time $T_\mathrm{cycle}$ of 1.69~ns. 
The maximum circuit latency for circulative paths, $\lambda$, is 100 clock cycles.

\begin{table*}[h]
	\caption{Detailed results and settings for K2000 and the first ten instances of G-set. 
	${\Delta_{\mathrm{TTS}}}$ and ${\Delta P_S}$ are the standard errors of TTS and $P_S$, respectively.
	They can be simply formulated as 
	${\Delta P_S} = \sqrt{(P_S-P_S^2)/N_{\mathrm{rep}}}$ 
	and 
	$\Delta_{\mathrm{TTS}} = \mathrm{TTS} \times {\Delta P_S} / [(1-P_S) \ln~(1-P_S)]$, 
	where $N_{\mathrm{rep}}$ is the number of repetitions for their evaluation~\cite{21}.}
	\includegraphics[width=16cm]{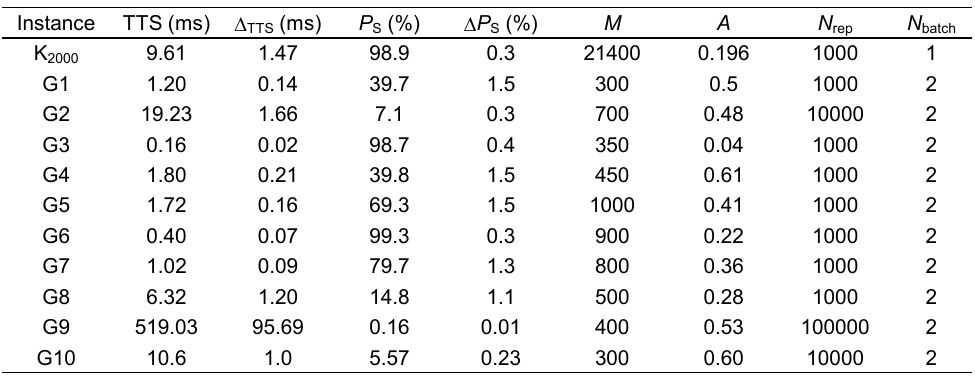}
	\label{table1}
\end{table*}

\begin{table*}[h]
	\caption{Detailed results and settings in the ascending order of TTSs for the 100 instances of the 700-spin Ising problem. The instance in Fig.~\ref{fig1} is highlighted in bold.
	${\Delta_{\mathrm{TTS}}}$ and ${\Delta P_S}$ are the standard errors of TTS and $P_S$, respectively 
	(see the caption of Table~\ref{table1} for their definitions).}
	\includegraphics[width=14cm]{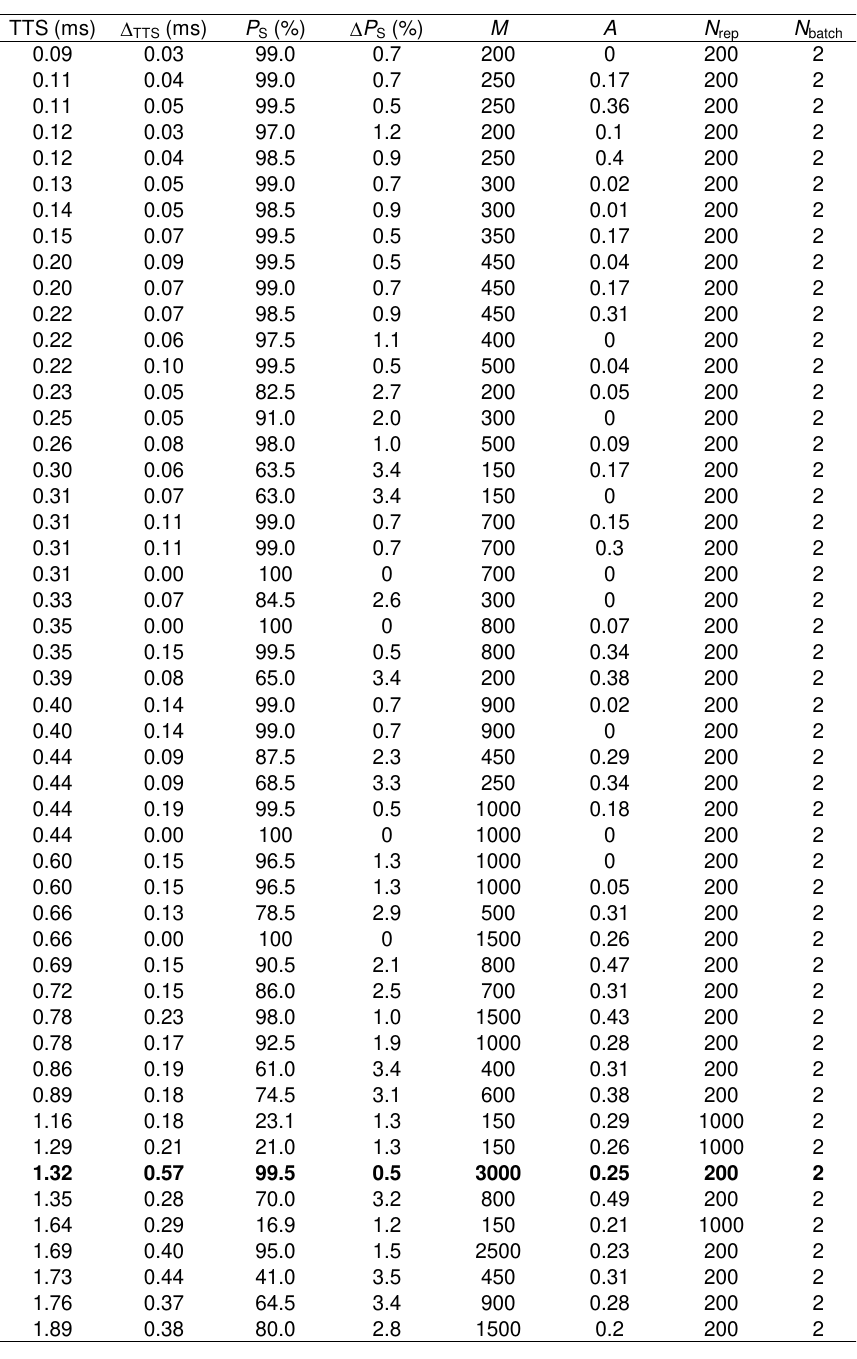}
	\label{table2}
\end{table*}

\begin{table*}[h]
	\includegraphics[width=14cm]{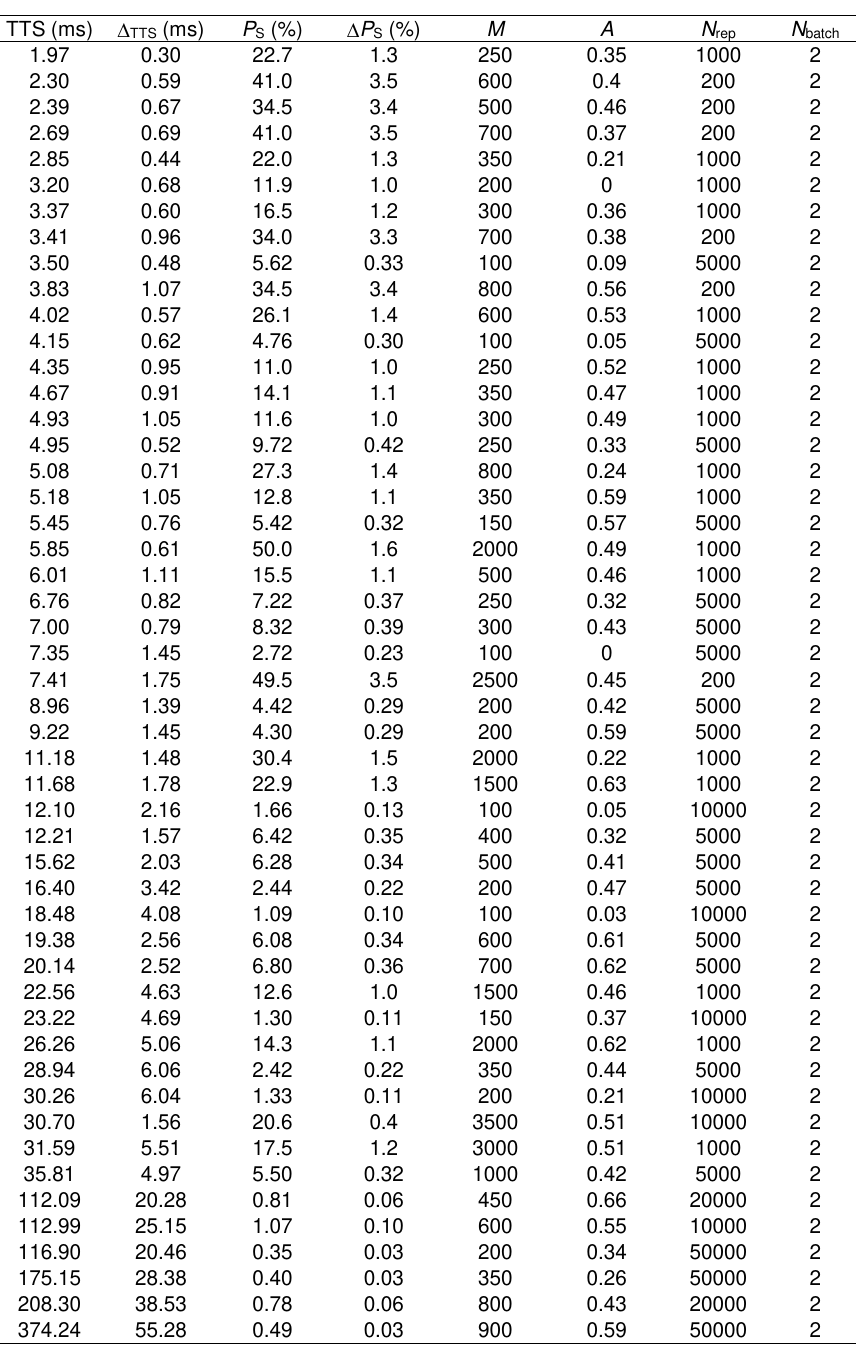}
\end{table*}

\begin{table*}[h]
	\caption{Implementation and performance of the GbSB-based FPGA machine.}
	\includegraphics[width=14cm]{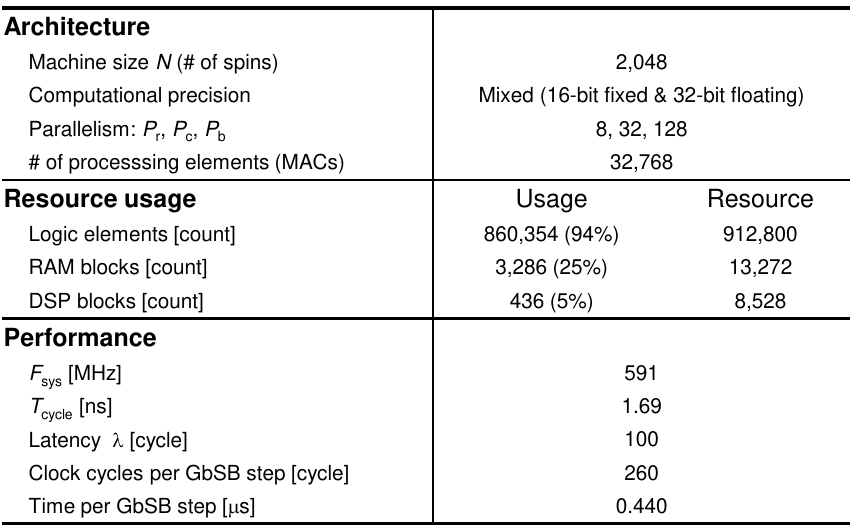}
	\label{table3}
\end{table*}

\end{appendix}

\end{document}